\documentclass[aps,prb,showpacs,groupedaddress,twocolumn]{revtex4}
\usepackage{amsmath,amssymb}
\usepackage{graphicx,color}

\newcommand{\Tr}{\text{Tr}}

\newcommand{\eavg}[1]{\left<{#1}\right>}
\newcommand{\abs}[1]{\left\lvert{#1}\right\rvert}

\DeclareMathOperator{\sgn}{sgn}
\newcommand{\integral}[1][]{\def\ArgI{{#1}}\IntegralRelay}
\newcommand\IntegralRelay[3][]{ \int_{\ArgI}^{#1} \mathrm{d}#2 \; #3 }
\newcommand{\mat}[1]{\begin{pmatrix}#1\end{pmatrix}}

\newcommand{\kG}{\check{G}}

\newcommand{\nh}{\hat{h}}

\newcommand{\nGA}{\hat{A}}
\newcommand{\nGR}{\hat{R}}

\newcommand{\nGK}{\hat{K}}
\newcommand{\nzero}{\hat{0}}

\newcommand{\ktau}{\check{\tau}}
\newcommand{\ntau}{\hat{\tau}}

\newcommand{\e}{\varepsilon}

\newcommand{\tk}{\ktau_K}

\newcommand{\xtermn}{\mathcal{T}_N}
\newcommand{\xterms}{\mathcal{T}_S}
\newcommand{\xterm}{\mathcal{T}}
\newcommand{\xnode}{\mathcal{N}}
\newcommand{\xall}{\mathcal{C}}

\begin{document}
\title{Circuit theory for noise in incoherent normal--superconducting dot structures}

\author{P.~Virtanen}
\email[]{Pauli.Virtanen@hut.fi}
\author{T.~T.~Heikkil\"a}
\email[]{Tero.T.Heikkila@hut.fi}
\affiliation{Low Temperature Laboratory, Helsinki University of
  Technology, P.O. Box 2200 FIN-02015 HUT, Finland.}

\begin{abstract}
We consider the current fluctuations in a mesoscopic circuit
consisting of nodes connected by arbitrary connectors, in a setup with
multiple normal or superconducting terminals. In the limit of weak
superconducting proximity effect, simplified equations for the
second-order cross-correlators can be derived from the general
counting field theory, and the result coincides with the semiclassical
principle of minimal correlations. We discuss the derivation of this
result in a multi-dot case.
\end{abstract}


\maketitle

Fluctuations of charge current in mesoscopic structures are in general
sensitive to the interactions and the fermionic nature of electrons.
In multi-terminal setups, the geometry of the circuit is important
for the cross-correlations, and in superconducting heterostructures,
also the Andreev reflection, the superconducting proximity effect and
transmission properties of NS interfaces need to be accounted for.

The general theory for the full counting statistics of current
fluctuations in multi-terminal structures was outlined in
Ref.~\onlinecite{nazarov02}.  The calculation of the second-order
correlators using this theory can be simplified, from complicated
$4\times4$ matrix equations to a Kirchoff-type system for scalar
parameters, using an approach discussed also, for example, in
Refs.~\onlinecite{samuelsson02,nagaev02}. In the incoherent case,
\cite{nagaev01,samuelsson02,belzig03} the result coincides with the
semiclassical principle of minimal
correlations. \cite{samuelsson02,nagaev02} In this paper we show the
derivation of this result in a multi-dot system, and consider a few
special cases.

The theory considers a network of normal ($\xtermn$) and
superconducting ($\xterms$) terminals ($\xterm=\xtermn\cup\xterms$)
and nodes ($\xnode$), connected by connectors.  Each connector $(i,j)$
is described by its transmission eigenvalues $T_n^{ij}$,
\cite{blanter00} and each node $j$ is characterized by a
Keldysh Green function $\kG_j$, which is a $4\times4$ matrix in the
Keldysh($\check{~}$) $\,\otimes\,$ Nambu($\hat{~}$) space.  In the
quasiclassical approximation, assuming stationary state and
isotropicity, these are only functions of energy, $\kG(\e)$.  

The statistics of the current in the circuit is connected to the
generating function $S(\{\chi_k\}_{k\in\xterm})$ of charge transfer,
which can be found by solving transport equations for the Green
functions.  In the stationary case at zero frequency, the noise
correlations $\tilde{S}_{kl}$ between the fluctuations $\delta
I_k=I_k-\eavg{I_k}$ of currents flowing into the terminals
$k,l\in\xterm$ relate to it through \cite{nazarov02,belzig02}
\begin{equation}\label{eq:noise}
  \tilde{S}_{kl} \equiv
  \int_{-\infty}^{\infty}\!\frac{\text{d}t}{2}\eavg{\left\{\delta I_k(t),\delta I_l(0)\right\}}
  = -\frac{e^2}{t_0} \left.\frac{\partial^2 S}{\partial\chi_k\partial\chi_l}\right\vert_{\{\chi_j\}=0}
  \,.
\end{equation}
Here, $t_0$ is the duration of the measurement, and the equality
applies provided this is much larger than the correlation time of the
fluctuations.

The boundary conditions for transport are assumed such that the
terminals are in an internal equilibrium, where the Green function has
the form
\begin{equation}\label{eq:Geq}
  \kG_{\text{eq}} = \mat{ \nGR & \nGK \\ \nzero & \nGA }, \;
  \begin{aligned}
  \nGR &= u\ntau_3+v i \ntau_2 \,, \;
  \nGK = \nGR \nh - \nh \nGA \,,
  \\
  \nGA &= -\ntau_3 \nGR^\dagger \ntau_3 \,,\; \nh = f_L + \ntau_3 f_T \,.
  \end{aligned}
\end{equation}
Here, $u=\abs{\e}/\sqrt{\e^2-\Delta^2}$, $v=\sgn(\e)\sqrt{u^2-1}$ are
the coherence factors, and $\Delta$ is the superconducting pair
amplitude. The functions $f_T(\e)=1-f_0(-\e)-f_0(\e)$ and
$f_L(\e)=f_0(-\e)-f_0(\e)$ are the symmetric and antisymmetric parts of
$f_0(\e)=[e^{(\e-eV)/(k_BT)}+1]^{-1}$, where $T$ is the temperature and
$V$ the potential of the terminal. We assume $V=0$ in all S terminals
to avoid time-dependent effects. For calculation of the statistics of
the current, the counting field theory additionally specifies the
rotation
\begin{equation}\label{eq:Gbc}
  \kG_k(\chi_k) = e^{i\chi_k \tk/2} \; \kG_{k,\text{eq}} \; e^{-i\chi\tk/2}
  \,,
  \quad
  \ktau_K\equiv\ktau_1\otimes\ntau_3
  \,,
\end{equation}
at each terminal $k$, which connects the ``counting fields'' $\chi_k$ to
the Green functions.

In circuit theory, \cite{nazarov99} transport is modeled by the
conservation of the matrix current at each node $i$
\begin{equation}\label{eq:currentconservation}
  \sum_{j\in\xall} \check{I}^{ij}=\check{0} \,, \;
  \check{I}^{ij}= \frac{2e^2}{\pi\hbar}\sum_n \frac{T_n^{ij} \left[\kG_j,\kG_i\right]}%
  {4 + T_n^{ij}\bigl(\left\{\kG_i,\kG_j\right\}-2\bigr)}
  \,.
\end{equation}
The sum runs over all nodes and terminals ($\xall=\xterm\cup\xnode$):
we assume the convention that $T_n^{ij}=0$ for $i=j$ and disconnected
points. This matrix is related to the observable charge and energy
currents by
\begin{equation}\label{eq:current}
  I^{ij} = \frac{1}{8 e} \integral[-\infty][\infty]{\e}{
    \Tr\left[\tk \check I^{ij} \right] } \,,
  \;
  I_E^{ij} = \frac{1}{8e^2} \integral[-\infty][\infty]{\e}{\e
    \Tr\left[\ktau_1\check I^{ij}\right] } \,.
\end{equation}
Their dependency on $\{\chi_i\}$, in turn, describes the generating
function of charge transfer: \cite{nazarov02}
\begin{equation}\label{eq:multiS}
  \mathrm{d}S(\{\chi_l\})
  = -\frac{t_0}{e} \sum_{k\in\xterm}
  \sum_{j\in\xall} I^{jk}(\{\chi_l\})\; \,\mathrm{d}(i\chi_k)
  \,.
\end{equation}
Determining the Green functions at the nodes from
Eqs.~(\ref{eq:Gbc},\ref{eq:currentconservation}) and finally applying
Eqs.~(\ref{eq:current},\ref{eq:multiS}), one can in principle find the
distribution of the fluctuations in the current. However, the problem
becomes considerably simpler if one is interested only in the second
moment of this distribution, i.e., the current noise as given in
Eq.~\eqref{eq:noise}.

We proceed calculating the noise by assuming that the superconducting
proximity effect is negligible, so that the anomalous parts
($\propto\ntau_1,\ntau_2)$ of the functions vanish in each
node.~\cite{samuelsson02,belzig03} Then, one can expand the Green
function at node $j$ to the first order in the counting fields
$\{\chi_k\}$, in the Nambu-diagonal form:
\cite{samuelsson02,nagaev02,houzet04}
\begin{equation}\label{eq:kelexpansion}
 \kG_j = \mat{ \ntau_3 & 2 \nh_j\ntau_3 \\ 0 & -\ntau_3 }
  + \sum_{k\in\xterm} i\chi_k \mat{
    -\nh_j \hat{b}^{j}_k & 4 \hat{c}^{j}_k-\hat{b}^j_k \\
     \hat{b}^{j}_k & \nh_j \hat{b}^{j}_k
  } + \ldots,
\end{equation}
where $\hat{b}_k^j(\e)=\hat{1}b_k^j(\abs{\e})$, $\hat{c}_k^j=c_T^{jk}+\ntau_3
c_L^{jk}$ and $\hat{h}=f_L+\ntau_3f_T$.  This satisfies the
quasiclassical normalization $\kG_j^2=\check{1}$ up to second order in
$\{\chi_k\}$. For the matrix currents, the above corresponds to the
expansion
\begin{equation}\label{eq:iexpansion}
  \check{I}^{ij}
  = \mat{\nzero & \hat{I}_0^{ij} \\ \nzero & \nzero}
  + \sum_{k\in\xterm} i\chi_k
  \mat{\nzero & \hat{I}_{c,k}^{ij}-\hat{I}_{b,k}^{ij} \\
       \hat{I}_{b,k}^{ij} & \nzero }
  + \ldots + \check{I}_{\text{coh.}}^{ij}
\end{equation}
of Eq.~\eqref{eq:currentconservation}, where $\hat{I}_0$,
$\hat{I}_{b,k}$ and $\hat{I}_{c,k}$ have the structure
$\hat{I}=\ntau_3 I(\ntau_3\e)$, due to symmetries in the Nambu
space. Here, $\check{I}_{\text{coh.}}(\{\chi_k\})$ contains the
off-diagonal Nambu-elements, present if $j$ corresponds to a
superconducting terminal. In what follows, we neglect this coherent
part of the current, assuming there are additional
decoherence-inducing sink terms in
Eq.~\eqref{eq:currentconservation}. \cite{samuelsson02,belzig03} This
is valid provided that the Thouless energy describing the inverse time
of flight through the node or the connector is much less than the
characteristic energy scales of the problem, or, if there is a strong
pair-breaking effect in the node, e.g., due to magnetic impurities.

One can then consider expansion~\eqref{eq:iexpansion} in detail,
assuming a node $i$ is connected to a node or terminal $j$. This
yields four independent equations of conservation:
\begin{equation}\label{eq:curcons}
  \sum_{j\in\xall} I_T^{ij} = 0 ,\;
  \sum_{j\in\xall} I_L^{ij} = 0 ,\; 
  \sum_{j\in\xall} I_{b,k}^{ij} = 0 ,\;
  \sum_{j\in\xall} I_{c,T,k}^{ij} = 0 \,,
\end{equation}
in which $I_T$ corresponds to the spectral charge current, $I_L$ to
the energy current, and the last two to a ``noise'' current, with the
symmetric part defined as
$I_{c,T,k}(\e)=I_{c,k}(\e)+I_{c,k}(-\e)$. The corresponding
antisymmetric current $I_{c,L,k}^{ij}$ is not needed, as we
concentrate on the noise in the charge current. The spectral
currents have the form
\begin{subequations}\label{eq:curdef}
\begin{gather}
  I_T^{ij} = g_{ij}(f_T^j - f_T^i) \,, \quad
  I_{b,k}^{ij} = g_{ij}(b^j_k - b_k^i) \,, \\
  I_L^{ij}=\begin{cases}
  0 & \text{for $j\in\xterms$ and $\abs{\e}<\abs{\Delta}$,} \\
  g_{ij}(f_L^j - f_L^i)  & \text{otherwise.}
  \end{cases}
\end{gather}
\end{subequations}
Thus, no energy current flows to the superconductors for
$\abs{\e}<\abs{\Delta}$. The fourth current is
\begin{equation}
  \frac{1}{4}I_{c,T,k}^{ij} = g_{ij}(c_T^{ik}-c_T^{jk}) 
  - (b_k^i-b_k^j)(s_{ij}(\e) + s_{ij}(-\e)) \,,
\end{equation}
but it can be eliminated, see below. 

The factors $g_{ij}$ and $s_{ij}(\e)$ appearing in the expansion can
be identified as the conductances and spectral noise densities
characteristic of the connectors, and their exact form depends on
whether the connector lies between two normal points (NN) or between a
normal and a superconducting point (NS).  The expressions for the NS
case are lengthy, so for simplicity we use here only the limits
$\e\ll\Delta$ and $\e\gg\Delta$ for superconducting Green's functions,
effectively neglecting the exact form of the superconducting density
of states (DOS). In this approximation, for an NS connector at
$\abs{\e}\gg\abs{\Delta}$ or an NN connector,
\begin{subequations}\label{eq:nnfanores}
\begin{gather}
  \begin{split}
    s_{ij}^{NN}(\e)
    = \frac{1}{4} g_{ij}^{NN}[
      & 2 - (f_L^i+f_T^i)^2 - (f_L^j+f_T^j)^2 \\
      & + F_{ij}^{NN} (f_L^i+f_T^i-f_L^j-f_T^j)^2 ] \,,
  \end{split} \label{eq:nnpectralnoise} \\
  g_{ij}^{NN} = \frac{e^2}{\pi\hbar}\sum_n T_n^{ij} \,, \;
  F_{ij}^{NN} = \frac{e^2}{g_{ij}^{NN} \pi\hbar}\sum_n T_n^{ij}(1-T_n^{ij}) \,.
\end{gather}
\end{subequations}
The result for an NS connector at $\abs{\e}\ll\abs{\Delta}$ is
\begin{subequations}\label{eq:nsfanores}
\begin{gather}
  s_{ij}^{NS}(\e)
  = \frac{1}{2} g_{ij}^{NS}[ 1 - (f_L^i)^2 - (f_T^i)^2 + F_{ij}^{NS}(f_T^i)^2 ] \,,
  \\
  g_{ij}^{NS} = \frac{e^2}{\pi\hbar}\sum_n \frac{2(T_n^{ij})^2}{(2-T_n^{ij})^2} \,,\\
  F_{ij}^{NS} = 
  \bigl(g_{ij}^{NS}\bigr)^{-1} \, \frac{e^2}{\pi\hbar}\sum_n \frac{16(T_n^{ij})^2}{(2-T_n^{ij})^4}(1-T_n^{ij}) \,,
\end{gather}
\end{subequations}
as found through an expansion of
Eq.~\eqref{eq:currentconservation}. Naturally, the results above agree
with expressions for the noise generated between two terminals, with
$F_{ij}$ being the differential Fano factor. \cite{blanter00,jong94}

The above equations are supplied with the boundary conditions
\begin{equation}\label{eq:parambc}
  b_k^l = \delta_{kl}\,, \quad c_k^l = 0\,, \quad 
  f_k(\e) = f_0(\e, V_k, T_k) \,,
\end{equation}
where $k$ and $l$ are indices of terminals. These can be found
by comparing expansion~\eqref{eq:kelexpansion} to Eq.~\eqref{eq:Gbc}
(for N terminals), and by examining the expression for $\check{I}$
(for S terminals).

Finally, Eqs.~(\ref{eq:noise},\ref{eq:current},\ref{eq:multiS}) yield
the result
\begin{align}
\tilde{S}_{kl}&=\sum_{j\in\xall}\integral[-\infty][\infty]{\e}{\frac{1}{8}I_{c,T,l}^{kj}}
\notag
\\
&= \sum_{(i,j)}
\integral[-\infty][\infty]{\e}{(b_k^i-b_k^j)(b_l^i-b_l^j) s_{ij}(\e)} \,, \label{eq:noisebx}
\,
\end{align}
for the correlations between terminals $k$ and $l$. In the last step,
we eliminated all $c_{T,k}^i$ from the set of equations, which
transforms the result to a sum over all connectors $(i,j)$ in the
circuit.

The equations above have a simple physical interpretation. The first
two of Eqs.~\eqref{eq:curcons} describe the conservation of charge (T)
and energy (L) currents at each energy interval
$[\e,\e+\mathrm{d}\e]$. With boundary
conditions~(\ref{eq:parambc},\ref{eq:nnfanores},\ref{eq:nsfanores}),
they yield distribution functions $f_L^i$, $f_T^i$ of electrons at the
nodes. In addition, one needs to solve from Eqs.~\eqref{eq:curcons}
the variable $b_k^i$, which characterizes the coupling between
terminal $k$ and node $i$. It turns out that this quantity is in fact
the characteristic potential introduced for semiclassical
multiterminal calculations. \cite{buettiker93} Knowing $f$, the
standard two-terminal relations \cite{beenakker97}
(\ref{eq:nnfanores},\ref{eq:nsfanores}) give the spectral noise
densities in each connector, and Eq.~\eqref{eq:noisebx} describes how
these couple to the terminals. The final result is similar to the
semiclassical result in diffusive metals, \cite{sukhorukov98}
and coincides with the result in dot systems, see below.

The assumption of all nodes being in the normal state resulted in a
simple way to handle superconductors in one special case: first, it
takes into account that no energy current enters superconductors at
$\abs{\e}<\abs{\Delta}$, and second, assumes that other effects due to
superconductivity are localized in only one connector, where both the
conductivity and the generated noise are modified. Our last
approximation of a piecewise constant superconducting DOS simplifies
the resulting expressions.

We implicitly assumed above that there is no inelastic scattering
which would drive the system towards equilibrium. However, following
Ref.~\onlinecite{nagaev95}, a strong relaxation of the distribution
function in a node may be modeled by assuming that $f_j$ has the form
of a Fermi function. In the case of relaxation due to strong
electron-electron scattering, the corresponding potential $V_j$ and
temperature $T_j$ can be determined by taking the two first moments,
$\int\text{d}\e$ and $\int\text{d}\e\e$ of Eqs.~\eqref{eq:curcons}:
\begin{subequations}\label{eq:volttempcons}
\begin{gather}
  \sum_{j\in\xall} g_{ij}(V_i - V_j) = 0\,, \quad \mathcal{L}^{-1}\equiv\frac{3e^2}{k_B^2\pi^2} \,, \label{eq:voltcons}\\
  \sum_{j\in\xall\setminus\xterms} g_{ij}[T_i^2 - T_j^2 + \mathcal{L}^{-1}(V_i^2 - V_j^2)] = 0 \,. \label{eq:tempcons}
\end{gather}
\end{subequations}
These describe the conservation of charge and energy currents. If some
of the nodes are in non-equilibrium, one can define the effective
voltages and temperatures
so that Eqs.~\eqref{eq:volttempcons} still apply for the whole
circuit.  In addition, one can model relaxation due to strong
electron-phonon coupling by forcing $T_i$ coincide with the lattice
temperature, so that only $V_i$ need to be determined.

It is illustrative to note that the quantum-mechanical counting-field
theory agrees with the well-known principle of minimal correlations,
which is often used in semiclassical
calculations. \cite{blanter00,samuelsson02} In a typical model, one
has the Langevin equations
\begin{equation}
  \sum_{j\in\xall} I^{ij} = 0 \,,\quad
  I^{ij} = g_{ij}(V_j - V_i) + \delta I^{ij} \,,
\end{equation}
where $\delta I^{ij}$ are the microscopic fluctuations of the current,
generated in the connector $(i,j)$. Eliminating voltages $V_i$ at the
nodes and assuming they do not fluctuate at the terminals, one finds
the result
\begin{equation}
  \delta I_k = \sum_{(i,j)}(b_k^i - b_k^j)\delta I^{ij} \,,
\end{equation}
for the fluctuations $\delta I_k$ in the current flowing to terminal
$k$. Assuming $\delta I^{ij}$ are independent and evaluating
$\frac{1}{2}\eavg{\{\delta I_k,\delta I_l\}}$, one finds
Eq.~\eqref{eq:noisebx}.  This coincides with the prediction from the
counting field theory, for an arbitrary circuit, provided it is
understood that $s_{ij}$ should be evaluated using the (average)
distribution functions at the nodes.  These may in general be in
non-equilibrium, and should be obtained from a kinetic
equation. Moreover, in the incoherent limit, the semiclassical result
is correct also in the presence of superconducting
terminals. \cite{nagaev01,samuelsson02}

The above discussion also clearly shows that an attempt to evaluate
the higher correlators of noise using the principle of minimal
correlations fails, as this corresponds to truncating
expansion~\eqref{eq:kelexpansion} after the first two terms. The
higher-order semiclassical corrections needed to fix this are
discussed for example in Ref.~\onlinecite{nagaev02}.

\begin{figure}\centering
  \includegraphics{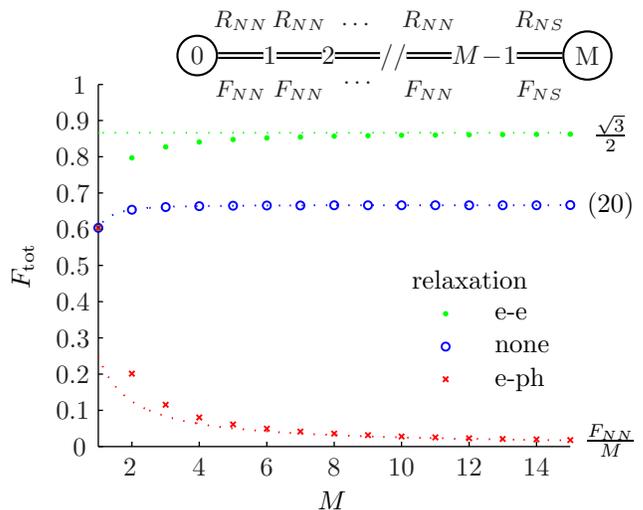}
  \caption{\label{fig:nodestring} (Color online). Total Fano factor
    $F_{\text{tot}}$ for a string of $N+1$ nodes (inset), connected by
    $M$ connectors with the transmission eigenvalue distribution
    $\rho(T)=2g(\pi\sqrt{T(1-T)})^{-1}$ of chaotic cavities. Results
    are shown for three types of relaxation in the nodes. Inset: $M$
    connectors in series.
  }
\end{figure}

Consider now an example setup that consists of $M-1$ nodes between two
terminals ``$0$'' and ``$M$'' (see inset of
Fig.~\ref{fig:nodestring}), and attempt to calculate its differential
Fano factor, $F_{\text{tot}}\equiv\frac{\partial\tilde{S}_{00}}{e
\partial I}\rvert_{I=0}$ at zero temperature $T_0=T_M=0$. For
simplicity of resulting expressions, we assume that all connectors are
identical, sharing the same distribution $\{T_n^{ij}\}$ of the
transmission eigenvalues.

First, if both terminals are normal, the application of
Eqs.~(\ref{eq:curcons},\ref{eq:curdef},\ref{eq:nnfanores},\ref{eq:parambc},\ref{eq:noisebx})
is analogous to the semiclassical calculation presented in
Ref.~\onlinecite{oberholzer02}, and yields the result
\begin{equation}\label{eq:stringfano}
  F_{\text{tot,NN}} = 
  \begin{cases}
    \frac{1}{3} + \frac{3F-1}{3M^2} \,, & \text{for no relaxation,} \\
    \frac{F}{M} \,, & \text{for $e$-ph relaxation,} \\
    \frac{\sqrt{3}}{4} \,, & \text{for $e$-$e$ relaxation, $M\rightarrow\infty$.}
  \end{cases}
\end{equation}
This shows that the limit $M\rightarrow\infty$ corresponds to the
diffusive limit, due to the isotropicity of electron momentum assumed
at the nodes.

If terminal $M$ is superconducting, and relaxation is negligible, we
need to apply
Eqs.~(\ref{eq:curcons},\ref{eq:curdef},\ref{eq:nnfanores},\ref{eq:nsfanores},\ref{eq:parambc},\ref{eq:noisebx}). In
this example $b_k^j$ are then straightforward to find, $f_L^j=f_L^0$
and $f_T^j=f_T^0 b_0^j + f_T^M b_M^j$ for $j=1,\ldots,M-1$.  Summation
in \eqref{eq:noisebx} then leads to a simple result
\begin{equation}
  F_{\text{tot}} 
  = \frac{2}{3} + 
  \frac{ (F_{NS} - \frac{2}{3})R_{NS}^{3} + \frac{M-1}{2}(F_{NN} - \frac{1}{3})R_{NN}^{3}}{[R_{NS} + (M-1)R_{NN}]^{3}} \,.
\end{equation}
Here $R_{NN}$, $F_{NN}$, $R_{NS}$ and $F_{NS}$ are the resistances and
differential Fano factors of the $NN$ and $NS$ connectors, as given in
Eqs.~\eqref{eq:nnfanores} and \eqref{eq:nsfanores}. The result applies
also for $M=1$, and in fact, for $M=2$ it is valid even if both
connectors have differing $\{T_n^{ij}\}$. In the limit
$M\rightarrow\infty$, the Fano factor again tends towards that of a
diffusive contact, showing the doubling of the shot
noise. \cite{jong94}

\begin{figure}
  \includegraphics{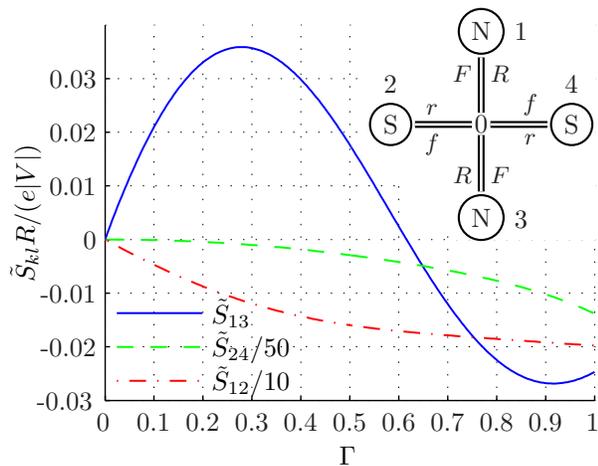}
  \caption{\label{fig:nnsscrosscorrelation} (Color online). Cross
    correlation $\tilde{S}_{13}$ between the two normal terminals
    changes sign as a function of contact transparency $\Gamma$, while
    $\tilde{S}_{24}$ and $\tilde{S}_{13}$ stay negative. Voltages are
    chosen $V_1=V_3=V$, $V_2=V_4=0$, and all connectors are assumed to
    be identical. Inset: A four-probe system. Terminals 1 and 3 are
    normal, 2 and 4 are superconducting. Connectors are assumed to
    have sub-gap resistances $R,r$ and differential Fano factors
    $F,f$.
  }
\end{figure}

Similar calculation shows that for strong inelastic e-ph scattering
one has
%
$F_{\text{tot,e-ph}}\sim F_{NN}/M$
%
for large $M$. For relaxation due to e-e scattering, in turn,
Eq.~\eqref{eq:tempcons} first gives the temperature profile $T_j =
[T_0^2+\mathcal{L}^{-1}(V_0^2-V_j^2)]^{1/2}$, where $V_j=b_0^j V_0$. From
Eqs.~(\ref{eq:noisebx},\ref{eq:nnpectralnoise}) one then finds
that
%
%
$F_{\text{tot,e-e}}=\sqrt{3}/2$ for $M\rightarrow\infty$, 
showing again the doubling of the noise. Numerical results for the
behavior at smaller $M$ are shown in Fig.~\ref{fig:nodestring}.

It is mostly straightforward to solve the current correlations in
multiterminal N-S systems, also discussed for example in
Refs.~\onlinecite{samuelsson02b,boerlin02,nagaev01nonloc}. For the
four-terminal setup shown in the inset of
Fig.~\ref{fig:nnsscrosscorrelation}, one obtains
\footnote{ This type of a problem can also be solved exactly, including
  the proximity effect, see Ref.~\onlinecite{boerlin02}.  }
\begin{align} \label{eq:crosscorr}
  \tilde{S}_{13}
  &= - c_1 e(\abs{V_1} + \abs{V_3} + \abs{V_1 + V_3}) - c_2 e\abs{V_1 - V_3}
  \\
  c_1 &= \frac{r}{16(r+R)^4}[rR(1\!+\!F)+2FR^2+2r^2(1\!-\!f)] \\
  c_2 &=
     \frac{r}{16(r+R)^4}\bigl[r(2r + R + 2r^2 R^{-1}) + 2r^2f \\
       &\qquad+ (2+rR^{-1})(2r^2+4rR+R^2)F \bigr] \,. \notag
\end{align}
Here, $R=R_{NN}$, $r=R_{NS}$, $F=F_{NN}$ and $f=F_{NS}$, and the
result is valid provided $\abs{V_i}\ll\abs{\Delta}$, $i=1,3$, and
$T=0$. If the connectors are assumed diffusive ($F=1/3$, $f=2/3$),
Eq.~\eqref{eq:crosscorr} agrees with Ref.~\onlinecite{nagaev01nonloc}.
One also finds that for $f>1$, the
cross-correlation~\eqref{eq:crosscorr} can be positive if $R$ is small
enough, \cite{samuelsson02b} contrary to the case in normal-state
circuits. For NS contacts with transparency $\Gamma$, this is
satisfied for $0<\Gamma<2(\sqrt{2}-1)$, as in
Refs.~\onlinecite{torres99,samuelsson02b}.  A different example, where
all four contacts are identical so that $R$ is not small, is shown in
Fig.~\ref{fig:nnsscrosscorrelation}.


In conclusion, we discuss a simple model for the transmission of noise
in multi-dot incoherent normal--superconducting structures, applying
the microscopic counting field theory. The formalism produces the
principle of minimal correlations, and has strong analogies with the
semiclassical theory of noise in diffusive structures.


We thank W.~Belzig for discussions, and P.~Samuelsson and
M.~B\"uttiker for pointing out their previous work in
Ref.~\onlinecite{samuelsson02}. TTH acknowledges the funding by the
Academy of Finland.

\end{document}